\documentclass[sigconf, nonacm]{acmart}
\acmConference[SoCC '24, Under review]{ACM Symposium on Cloud Computing}{November 20--22, 2024}{Redmond, WA, USA}
\acmISBN{978-1-4503-XXXX-X/18/06}

\usepackage{fontawesome}
\usepackage{enumitem}
\usepackage{tikz}
\usepackage{pifont}
\usepackage[capitalize,noabbrev]{cleveref}
\usepackage{listings}
\usepackage{fancyvrb}
\usepackage{balance}

\newcommand{\ourbench}[0]{\textsc{AIOpsLab}}

\definecolor{edgeblue}{RGB}{0, 0, 200}
\definecolor{edgegreen}{RGB}{0, 200, 0}
\definecolor{gptgreen}{RGB}{0, 166, 126}
\definecolor{scholarpurple}{RGB}{169, 1, 251}
\definecolor{bgcode}{rgb}{0.95,0.95,0.95}
\definecolor{githubgreen}{rgb}{0.564, 0.933, 0.564}
\definecolor{orange}{rgb}{1,0.5,0}

\definecolor{codegreen}{rgb}{0,0.6,0}
\definecolor{codegray}{rgb}{0.5,0.5,0.5}
\definecolor{backcolour}{RGB}{245,248,250}
\definecolor{emph}{RGB}{166,88,53}
\definecolor{nightblue}{RGB}{9,49,105}
\definecolor{keywords}{RGB}{207,33,46}
\definecolor{lightpurple}{RGB}{130,81,223}
\definecolor{examplebg}{RGB}{250,243,240}
\definecolor{codemph}{RGB}{150,30,30}
\newcommand{\Paragraph}[1]{\smallskip\noindent{\bf #1.}}

\newcommand*\circled[1]{\tikz[baseline=(char.base)]{
            \node[shape=circle,draw,inner sep=0.25pt] (char) {#1};}}

\newcommand{\HighlightPara}[1]{\smallskip\noindent{\bf \inlinebox{#1.}}}

\usepackage{xcolor}

\definecolor{mscolor}{rgb}{0.1,0.1,0.9}

\definecolor{gscolor}{rgb}{0.1,0.8,0.1}

\definecolor{yscolor}{rgb}{0.7,0.3,0.7}

\definecolor{mhcolor}{rgb}{0,0.9,0.9}

\definecolor{jmcolor}{rgb}{0.2,0.9,0.9}

\definecolor{pbcolor}{rgb}{0.5,0.5,0.5}
\newcommand{\pb}[1]{}

\definecolor{owncolor}{rgb}{0.7,0.7,0.7}
\newcommand{\own}[1]{}

\definecolor{yfcolor}{rgb}{1,0,0}

\definecolor{codegreen}{rgb}{0,0.45,0}
\definecolor{codegray}{rgb}{0.5,0.5,0.5}
\definecolor{codepurple}{rgb}{0.58,0,0.82}
\definecolor{backcolour}{rgb}{0.95,0.95,0.92}
\definecolor{mauve}{rgb}{0.58,0,0.82}

\newcommand{\code}[1][]{\lstinline[basicstyle=\small\ttfamily,#1]}
\newcommand{\codetiny}[1][]{\lstinline[language=bash,basicstyle=\scriptsize\ttfamily,#1]}
\newcommand{\codefoot}[1][]{\lstinline[basicstyle=\footnotesize\ttfamily,#1]}

\lstdefinestyle{mystyle}{
    commentstyle=\color{codegreen},
    keywordstyle=\color{keywords},
    stringstyle=\color{nightblue},
    basicstyle=\ttfamily\footnotesize,
    breakatwhitespace=false,
    breaklines=true,
    captionpos=b,
    keepspaces=true,
    numberstyle=\tiny\color{codegray},
    numbers=none,
    numbersep=1pt,
    showspaces=false,
    showstringspaces=false,
    showtabs=false,
    tabsize=2,
    frame=tb,
    emph={Suggest, Assert},
    emphstyle={\color{purple}},
    emphstyle={[2]\color{codemph}},
    escapeinside={(*@}{@*)},
    keywords={ValueErrors, TypeErrors, DataFormatErrors, AssertionErrors, EnvironmentErrors, indexer, _parse_sigmf, SignalDesc, SignalCap}
}

\lstMakeShortInline[keywordstyle=,%
                    flexiblecolumns=false,%
                    mathescape=false,%
                    basicstyle=\ttfamily]@

\usepackage[most]{tcolorbox}
\usepackage{lipsum}
\usepackage{tocloft}
\usepackage{etoolbox}
\usepackage{mdframed}
\usepackage{xcolor}

\newcommand{\agentbox}[2]{
  \begin{tcolorbox}[colback=blue!5!white,colframe=blue!75!black,boxsep=0.5pt,title=Agent]
    \footnotesize{\textbf{Thought:} #1}\vspace{1.25pt}\\#2
  \end{tcolorbox}
}

\newcommand{\obsbox}[1]{
  \begin{tcolorbox}[colback=green!5!white,colframe=green!45!black,boxsep=0.5pt,title=Observation]
    \footnotesize{#1}
  \end{tcolorbox}
}

\newcommand{\orchbox}[1]{
  \begin{tcolorbox}[colback=gray!5!white,colframe=gray!75!black,boxsep=0.5pt,title=Orchestrator]
    \footnotesize{#1}
  \end{tcolorbox}
}

\newtcbox{\inlinebox}[1][]{enhanced, box align=base, nobeforeafter, colback=gray!10, colframe=gray!50, boxrule=0.5pt, arc=4pt, left=1pt, right=1pt, top=-1pt, bottom=-1pt, #1}

\AtBeginDocument{%
  }

\begin{document}

% \title{Towards a Principled Framework for Evaluating AI Agents in Operational Resilience of Cloud Services}

% @chetan:
\title{Building AI Agents for Autonomous Clouds: \\Challenges and Design Principles} 

\subtitle{Vision Paper}

% Paper type: As a subtitle, papers should indicate the submission type (e.g., research, industry, vision, work-in-progress).

% * Authors: Chetan Bansal (Microsoft);
%  Dax Vandevoorde (Microsoft Research);
%  Gagan Somashekar (Microsoft);
%  Jonathan Mace (Microsoft Research);
%  Manish Shetty (Microsoft);
%  Minghua Ma (Microsoft);
%  Pedro Las-Casas (Microsoft Research);
%  Saravan Rajmohan (Microsoft);
%  Shachee Mishra Gupta (Microsoft);
%  Suman Nath (Microsoft Research);
%  Xuchao Zhang (Microsoft);
%  Yinfang Chen (Microsoft);
%  Yogesh Simmhan (Microsoft)

\author{Manish Shetty$^{1,2}$, Yinfang Chen$^{1,3}$, Gagan Somashekar$^1$, Minghua Ma$^1$, Yogesh Simmhan$^{1,4}$,\\
Xuchao Zhang$^1$, Jonathan Mace$^5$, Dax Vandevoorde$^{5,6}$, Pedro Las-Casas$^1$, Shachee Mishra Gupta$^1$, Suman Nath$^5$, Chetan Bansal$^1$, Saravan Rajmohan$^1$}
\affiliation{
  \institution{$^1$Microsoft, $^2$University of California, Berkeley, $^3$University of Illinois Urbana-Champaign, $^4$Indian Institue of Science, $^5$Microsoft Research, $^6$Agnes Scott College}
  \country{}
}

\begin{abstract}
% \ys{the title leans towards designing agents for the cloud, while the paper is actually on agent evaluation frameworks to achieve this indirectly...recheck pitch in intro for broader claim...done}

The rapid growth in the use of Large Language Models (LLMs) and AI Agents as part of software development and deployment is revolutionizing the information technology landscape.
While code generation receives significant attention, a higher-impact application lies in using AI agents for operational resilience of cloud services, which currently require significant human effort and domain knowledge. 
%While a major focus has been on using AI agents for code generation, of less prominence but higher-impact area is using such AI agents for operational resilience of cloud services, which requires significant human effort and domain knowledge.
%
There is a growing interest in AI for IT Operations (AIOps) which aims to automate complex operational tasks, like fault localization and root cause analysis, thereby reducing human intervention and customer impact. However, achieving the vision of autonomous and self-healing clouds
% ~\cite{6257956,10.1007/978-3-642-10665-1_5} 
through AIOps is hampered by the lack of standardized frameworks for building, evaluating, and improving AIOps agents.
%There is a growing interest in the realm of AIOps in exploring the use of AI agents to automate operations and resilience engineers during fault identification and mitigation, and reducing the time to system or service restoration. However, there is a lack of principled means to evaluate the efficacy of such AIOps agents.
%
This vision paper lays the groundwork for such a framework by first framing the requirements and then discussing design decisions that satisfy them.
%In this vision paper, we lay out the requirements for an evaluation harness to benchmark AIOps and traditional tools for operational management and resilience. 
We also propose \ourbench{}, a prototype implementation leveraging agent-cloud-interface that orchestrates an application, injects real-time faults using chaos engineering, and interfaces with an agent to localize and resolve the faults. We report promising results and lay the groundwork to build a modular and robust %evaluation platform for AIOps.
framework for building, evaluating, and improving agents for autonomous clouds.
\end{abstract}


\maketitle

%% Uncomment this to enable page numbering. OMit this before submission.
\thispagestyle{plain}
\pagestyle{plain}

% Vision papers is dual anonymous (i.e., author identities will be kept confidential from reviewers during the review process, and vice versa). Authors must make a good faith effort to anonymize their submissions, and they should not identify themselves either explicitly or by implication (e.g., through the references or acknowledgments).
% your submission must use an anonymized system/project name that is different from any used in such contexts.

\section{Introduction}
\pb{1.25pgs}\own{YS}
% [YS/GS/MM]

%% Clouds are complex
IT applications and services for enterprises and web-scale companies are becoming more complex to develop, deploy, and maintain. The broad adoption of the micro-services pattern to compose complex applications and serverless deployment models hosted on the cloud has arguably simplified application development and scaling. But this has also introduced more moving parts that make their operations, reliability, fault diagnosis and recovery even harder~\cite{8497007}. 
%
%
%% This leads to faults which have a business impact
Cloud services have set the expectation of five-9s of availability, and missing it can affect customer satisfaction. At the same time, failures in the clouds are becoming more frequent and with
% In real-world operational environments, system failures can have 
significant impact~\cite{amazonoutage, awsoutage, msoutage}. 
For instance, for one major recent outage, the estimated cost for one hour of service downtime for Amazon was approximately \$100 million~\cite{amazonoutage}.

%% Complexity makes human effort inadequate
Site Reliability Engineers (SRE) and DevOps engineers are tasked with deployment and operational maintenance of cloud services~\cite{hybrid-cloud}. They typically follow four steps when rectifying faults: fault detection, triaging, localization, and mitigation. This scope can be broader when we include proactive fault predictions and other preemptive maintenance. Many tools help with monitoring, anomaly detection, incident triage, root cause analysis, etc. \cite{bansal2020decaf, yan2023aegis}. However, SREs face information overload and complex decision-making in operating large-scale cloud environments. 
% Ghosh et al. \cite{ghosh2022fight} did 
E.g., a large-scale study of outages in Microsoft Teams found that root-causing and mitigation require significant manual effort, context awareness and technical expertise~\cite{ghosh2022fight}.
% E.g., in CompanyX \textit{(anonymized)}, one of the largest global cloud service providers and where the authors are from, there were \review{XYZ incidents} within a \review{XYZ month period} in one of the product portfolios, of which \review{XYZ} had the highest level of severity requiring resolution within \review{XYZ hours}.
%
Hence, the need for \textit{AIOps Agents}, which can automatically detect, localize and mitigate faults with minimal human intervention, is becoming critical. These AIOps agents are key to achieve the vision of Autonomous Clouds.

%% Need for AI tools.
% in the face of frequent and impactful outages~\cite{amazonoutage, awsoutage, msoutage}. 
% 
While the concept of self-healing clouds has been proposed earlier~\cite{6257956,10.1007/978-3-642-10665-1_5}, the emerging adoption of \textit{AIOps}, i.e., the use of AI to support IT Operations, is making this a reality~\cite{zhao2023robust, he2022empirical, ma2018robust, bmeg, TraceArk, xie2023unsupervised, Zhang:2018,ganatra2023detection, gamma14,li2021practical}. Here, AI algorithms and agents leverage the monitoring infrastructure to rapidly and efficiently track the health, identify failures and mitigate them within the operational environment~\cite{notaro2021survey}.
More recently, agents are able to converse with Large Language Models (LLMs) using an observe--thought--action pattern to localize problems, explore possible causes, and enact solutions to achieve recovery in a semi-autonomous manner~\cite{sweagent, panda-cidr,DBLP:conf/eurosys/ChenXMKGSCGFWZG24, DBLP:journals/corr/abs-2401-13810, jin2023assess}. 
% These tools meaningfully assist in fault diagnosis and recovery using an .
% 
We are at the cusp of AIOps agents being able to independently carry out these tasks in a matter of minutes within production environments, compared to several hours taken even by experts~\cite{ma2020diagnosing,wang2023rcagent}.
% 
% NOTE: The old version of the above text is below. Feel free to revert/merge:
% 
% There is a growing interest in the adoption of AIOps~\cite{zhao2023robust,he2022empirical,ma2018robust,TraceArk,xie2023unsupervised,Zhang:2018,ganatra2023detection, gamma14,li2021practical}, i.e., the use of AI to support IT Operations. This has been heightened through recent works that demonstrate the use of agents guided by Large Language Models (LLMs) to meaningfully assist in fault diagnosis and recovery using an observe--thought--action pattern~\cite{sweagent,DBLP:conf/eurosys/ChenXMKGSCGFWZG24,DBLP:journals/corr/abs-2401-13810, jin2023assess}. Such accelerated advances can lead to AIOps agents in the near future semi-autonomously or fully independently carry out many of these tasks in a matter of minutes, compared to several hours taken even by experts~\cite{ma2020diagnosing}.
% \ys{move come of the citations to related work to save space?}

The design, development, evaluation and iterative improvement of AIOps agents are challenging. While the adoption of AI has seen rapid progress for coding and programming~\cite{sweagent, aider} due to the availability of platforms like WebArena~\cite{webarena}, R2E~\cite{r2e} and benchmarks like HumanEval\cite{humaneval}, LiveCodeBench~\cite{lcb}, SWE-bench~\cite{swebench}, the same is lacking for cloud operations. 
Existing tools address individual aspects of the AIOps lifecycle: observability and introspection tools~\cite{10.1145/3611643.3613881,simonsson2021observability}, application suites~\cite{deathstarbench,xfbench-2024}, chaos engineering and fault injection tools~\cite{chaosmonkey,chaosblade,chaosmesh}, agent-computer interfaces~\cite{sweagent}, etc., but do not cohesively integrate them.

Moreover, recent prior works on leveraging AIOps agents for cloud operations like RCAgent~\cite{wang2023rcagent}, and Zhang et al.~\cite{rcs-fse-2024} use proprietary services and datasets. Other prior works use frameworks specific to the solutions that they are building~\cite{gamma14}, or \textit{ad hoc} and static benchmarks and metrics~\cite{opseval} that fail to capture the dynamic nature of real-world cloud services. Furthermore, current approaches do not agree on standard metrics or a standard taxonomy for operational tasks.
\textit{This calls for a standardised and principled framework for building, testing, comparing and improving AIOps agents}. %Specifically, we need to frame a set of common principles to design such an AIOps framework which can simplify and accelerate the development and adoption of AIOps agents to achieve autonomy in cloud operations.
The framework should allow agents to interact with realistic service operation tasks in a reproducible manner. It must be flexible in extending to new applications, workloads, and faults. Importantly, it should go beyond just evaluating the AI agents and also enable users to improve the agents themselves, e.g.,
by providing sufficient observability and even serving as a training environment (``gym'') to generate samples to learn on~\cite{rcs-fse-2024}.

\smallskip
In this vision paper, we draw upon our experiences from developing AIOps agents at Microsoft 
% \textit{(anonymized)}, one of the largest global cloud service providers, 
to make these contributions:
\begin{enumerate}[leftmargin=*, itemsep=1em]
\item We envision the requirements for a holistic framework to enable the design, development, evaluation, and enhancement of AIOps agents that, additionally, serves the purpose of reproducible, standardized, interoperable and scalable benchmarks. We also suggest key design decisions to build it (\S~\ref{sec:vision}).

\item We describe a prototype framework, \ourbench{}, that adopts this design to combine workload and fault generators to mimic production incidents and an agent-cloud interface for orchestrating the service operation lifecycle (\S~\ref{sec:arch}). This is the foundation for our ongoing work on a more comprehensive framework.

\item We report a case study on using this preliminary framework to evaluate an LLM agent for two critical operations tasks: Fault Detection and Mitigation (\S~\ref{sec:eval}).
\end{enumerate}

\section{Background and Related Work}
\label{sec:related}
% \pb{0.5pgs}
%\own{MH/YC}
% \mh{for tasks, we may have detection, localization, triage, diagnosis, mitigation, resolution}
% "classic" fault management tools: Nehza, Pinpoint
% Older AIOps tools: TIST survey
% LLM tools: SWE-Agent, PandaDB, RSA-FSE-20024

Next, we identify gaps in existing work for evaluating AIOps tools and agents in a standardized, realistic and reliable manner.

\Paragraph{Fault Mitigation Lifecycle} A typical incident goes through four stages: (1) \emph{Detection} \cite{ma2021jump,zeng2021watson,zeng2022shadewatcher}: When an anomalous system behaviour is observed, an alert is raised by monitors or users of the service (internal engineers or external customers) and reported to the Incident Management System (IcM). (2) \emph{Triaging} \cite{bansal2020decaf, chen2019empirical, chen2019continuous}: After the detection, the incident is notified to the On-Call Engineers (OCEs) to begin investigation and then assigned to the most appropriate engineering team. (3) \emph{Diagnosis} \cite{ma2020diagnosing, DBLP:conf/eurosys/ChenXMKGSCGFWZG24, DBLP:journals/corr/abs-2401-13810, deepanalyze}: The assigned engineers inspect different aspects of the incident and have several rounds of back-and-forth communication to identify the root cause. (4) \emph{Mitigation} \cite{jiang2020mitigate, ahmed2023recommending, shetty2022autotsg}: Several actions are taken by the team to mitigate the incident and to restore service health. The outcome is also updated postmortem in the IcM.
% \ys{this is repetitive from \S~\ref{sec:example}. Fold this above?}

% OpsEval, AgentOps-AI/agentops
%\own{GS/MS}
% ADBench/AnomalyBench - only single task (detection) - no env interaction
% LogParser - single task (incident parsing) - no env interaction
% OpsEval - Ops QA - no environment interaction
% AgentOps and AgentBench - generic agent benchmarks

%\Paragraph{AIOps Agents}
%\ys{Gagan: SWE-Agent~\cite{sweagent}, Panda~\cite{panda-cidr}, RSA-FSE-2024~\cite{rcs-fse-2024} and Rcagent~\cite{wang2023rcagent} are good papers to briefly cover here as examples of aiops agents, e.g., you can mention in-context learning, using prior data for prompt engineering~\cite{rcs-fse-2024}; retrofitting human interfaces for agents~\cite{sweagent}; giving a chance for humans to approve before database changes and to explain possible side-effects~\cite{panda-cidr};}

\Paragraph{AIOps Benchmarks}
Several benchmarks have been proposed to evaluate AIOps at various stages of the software lifecycle. For instance, \textsc{ADBench}~\cite{adbench} and \textsc{AnomalyBench}~\cite{anomalybench} evaluate anomaly detection, and \textsc{LogParser}~\cite{logparser} evaluates log parsing. However, these benchmarks focus only on single aspects of incident management.
More recently, LLMs have powered a new wave of AIOps tools and autonomous agents, forcing evaluations to evolve. General benchmarks~\cite{agentops, agentbench} like \textsc{AgentBench} task LLMs to operate as autonomous agents in diverse environments, solving problems across domains like operating systems and databases.

That said, specialized AIOps evaluations have received less attention. \textsc{OpsEval} \cite{opseval} attempts to address this with a question-answer-based evaluation for LLMs, but it disconnects from real-world operational challenges that require complex debugging, code understanding, and multi-step fault resolution.
Our proposed work on \ourbench{} bridges this gap. We believe using real services, workloads, and faults to create problems and enable executing concrete actions (e.g., run commands) is necessary for the reliable evaluation of state-of-the-art AIOps solutions, and to enhance them.

% DeathStar, XFBench
%\own{GS/MS}

\Paragraph{Application Benchmark Suites} Relevant to this work are benchmark suites for cloud applications \cite{ferdman2012clearing, hauswald2015sirius, kasture2016tailbench, wang2014bigdatabench,xfbench-2024}. \citet{ferdman2012clearing} presented \texttt{Cloudsuite} to study the architectural implications, and TailBench \cite{kasture2016tailbench} proposes a methodology to analyze the performance of web servers and database services. Further, the emergence of microservices has prompted recent work to study their characteristics and requirements \cite{kratzke2016ppbench, sriraman2018mu, ueda2016workload, zhou2018benchmarking}. The popular DeathstarBench \cite{deathstarbench} differentiates from these studies by focusing on diverse large-scale applications with tens of unique microservices, allowing studying effects that only emerge at scale, such as network contention and cascading Quality of Service (QoS) violations due to dependencies between tiers.
Beyond static application suites~\cite{deathstarbench, trainticket}, \textsc{BluePrint}~\cite{blueprint}
provides the ability to reconfigure applications and iteratively generate variants. We integrate with such 
% These complement for \ourbench{} sources of 
application workloads to enable benchmarking of AIOps solutions.

%\own{MH/YC}
% 3MileBeach, microFi, 

\Paragraph{Chaos Engineering and Fault Injection} Prior work developed fault injection techniques aimed at applications and distributed backend systems, including storage and data processing \cite{marinescu2009lfi,Banabic2012,Christakis2017a,zhang2012amplifying,jepsen, Pillai2014,alquraan2018analysis,lu2019crashtuner, chen2020cofi, leesatapornwongsa2014samc, gunawi2011fate, majumdar2018why, Ju2013, heorhiadi2016gremlin,alagappan2016correlated,Mohan2018,sun:osdi:22,canini2012nice}.
However, these existing techniques fall short in providing a generic, one-click fault generator that can be universally applied across various microservices. The limitations are multifaceted: many rely on application or domain-specific knowledge to create policies and oracles, making them unsuitable for the diverse requirements of AIOps \cite{lu2019crashtuner, chen2020cofi, leesatapornwongsa2014samc,Mohan2018,sun:osdi:22,gu:sosp:23}.
Others offer mechanisms without automated policies or oracles beyond simple crashes, requiring developers to manually implement complex functionalities \cite{meiklejohn2021,heorhiadi2016gremlin,Pillai2014,alquraan2018analysis,Christakis2017a,fis,Hunt1999,Ju2013,gunawi2011fate,Reynolds2006,Marinescu2010}.
Additionally, fault injections at a single level (e.g., HTTP) do not adequately expose root-causes of failures due to: 1) their coarse-grained nature, 2) challenge of constructing meaningful error objects, and 3) a lack of consideration for dependencies between microservices \cite{envoy,istio,marinescu2009lfi,Marinescu2011efficient,Christakis2017a,zhang2012amplifying,Jiang2020}.
\begin{figure*}[!t]
  \centering
  \includegraphics[trim={3.5cm 3.75cm 3.5cm 3.75cm},clip,width=2\columnwidth]{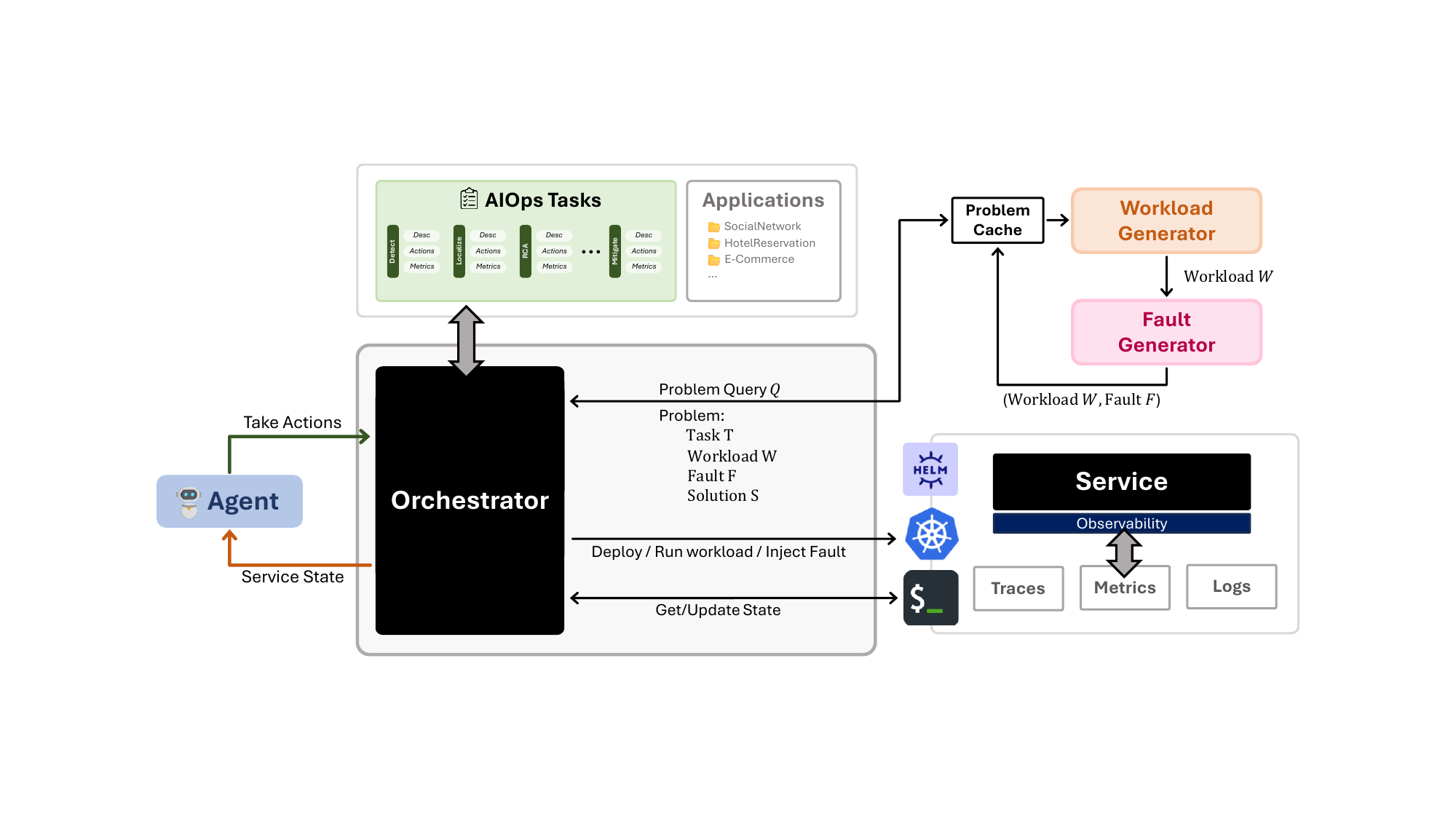}
  % \vspace{-25pt}
  \caption{\textbf{System Architecture of \ourbench{}.} \textmd{The Orchestrator coordinates interactions between various system elements and serves as the Agent-Cloud-Interface (ACI). 
  Agents engage with the Orchestrator to solve tasks, receiving a problem description, instructions, and relevant APIs. The Orchestrator generates diverse problems using the Workload and Fault Generators, injecting these into applications it can deploy.
  The deployed service has observability at multiple layers, providing telemetry, traces, and logs. The Orchestrator communicates with the service and the cloud using several tools such as \texttt{Kubernetes}, \texttt{Helm}, and even a \texttt{Shell}.
  Agents act via the Orchestrator, which executes them and updates the service's state. The Orchestrator evaluates the final solution using predefined metrics for the task.
  }}
  \label{fig:overview}
  %\vspace{-0.05in}
\end{figure*}

\section{\NoCaseChange{\ourbench{}}: A Principled Vision}
\label{sec:vision}
\pb{2.5pgs}
In this section, we first discuss the principles that we believe should guide the envisioned framework (\ref{sec:requirements}), followed by a discussion of the design choices that can lead to such a framework (\ref{sec:design}).

\subsection{Requirements and Principles}
\own{YS}\pb{1pg}
\label{sec:requirements}
% 
%\gs{I think we could stress in R5 that the applications/services are live/deployed}
% 
% \ys{need to trim down any verbose description. Suggestions welcome.}
%\ms{took a pass and trimmed descriptions; also made headers concise, feel free to revert.}

\begin{enumerate}[start=1,label={\itshape (\bfseries R\arabic*)},leftmargin=*,itemindent=20pt,itemsep=4pt]
    \item \label{req:1}
    \textbf{\textit{Modular Design for applications, workloads, and agents.}}
    % \textit{The design needs to be modular, allowing integration of new application benchmarks and workloads, reliability engineering workloads, and recovery agents.}
    % 
    An effective evaluation system should seamlessly incorporate existing and evolving resources on application workloads, fault injection models, and resilience techniques for fault detection and resolution. It must be flexible, supporting easy integration of new components through standard interfaces. This allows for varied use cases, including enterprise and end-user assessments of fault and recovery strategies, AIOps research on new agents using established error models and workloads, and red-team simulations of complex faults on synthetic applications to test mitigation effectiveness.
     % (\S~\ref{sec:related})

    \item \label{req:2} \textbf{\textit{Flexible Interface for human, digital, and AI agents.}}
    % \textit{Human, digital, and AI agents should be able to interface with the system to explore fault injection and mitigation strategies. In particular, the interfaces exposed should be amenable to LLM-based conversational agents.}
    % 
    The framework must provide diverse interfaces to agents, spanning the unique requirements of humans to LLM-based agents. Humans might use a web interface for log review and command execution, digital agents need APIs for integration, and conversational LLM agents require prompt-based interaction for requests and responses.

    \item \label{req:3} \textbf{\textit{Scalability at various operational and temporal scales.}}
    % \textit{The framework should operate at small (single VM) and large (100s of VMs) scales, and at short (minutes) and long (days) time scales.}
    % 
    Operating at diverse spatial (single VM to clusters) and temporal scales (minutes to days) is paramount to make the framework amenable to different use cases and resource availability, e.g., from large enterprises with complex deployments and a wide fault surface to software engineering course assignment with a tiny scenario. Operating across time scales also allows faults that gradually emerge (e.g., memory leaks) or occur periodically to be detected, predicted, and preemptively mitigated.

    \item \label{req:4} \textbf{\textit{Reproducible Setup for reliable measurements.}}
    % \textit{The design should enable reproducibility of the setup and, preferably, the results. There should be sufficient visibility to quantitatively and qualitatively measure the effectiveness of the mitigation strategy, allowing different evaluation metrics to be incorporated.}
    % 
    % An essential aspect of 
    An evaluation framework needs consistent and, ideally, automated deployments for reproducible and standardized assessment of mitigation strategies. Challenges may arise from non-deterministic elements in applications or faults, which should stem from external models, not the framework itself.
    While evaluation metrics can differ depending on the user or context, the framework should offer default objective metrics (e.g., accuracy, time to mitigate) and support the creation of more intricate measures with the data it provides.

    \item \label{req:5} \textbf{\textit{Versatility in operating environments.}}
    % 
    % \textit{It should be possible to operate the system in production, sandboxed and simulated environments.}
    % 
    The operating environments may vary based on the needs of the user. Chaos engineering~\cite{chaosmonkey}, for instance, advocates for fault injections directly in production environments, necessitating a framework capable of integrating with live applications. Alternatively, a sandboxed deployment with a simpler variant of production applications may be preferred. Synthetic and emulated systems~\cite{violet:tcps,calheiros2011cloudsim} can also help evaluate large-scale what-if scenarios.

    \item \label{req:6} \textbf{\textit{Comprehensive and Cross-Layer fault support.}}
    % 
    % \textit{The framework should allow faults and their mitigation to be introduced and resolved at different layers of the stack. The system should enable complex faults interlinked faults to be introduced and diagnosed.} 
    % 
    The framework should enable the introduction of faults across the entire stack, including hardware, network, OS, middleware, application, and external services. It must offer capabilities to simulate realistic faults inspired by real-world production incidents, which can cause cascading effects across distributed components, as well as synthetic faults for assessing potential future scenarios.

    \item\label{req:7} \textbf{\textit{Diverse and realistic workload conditions.}}
    % 
    % \textit{The framework should allow for generating diverse and complex workload conditions.}
    % 
    Workloads across domains have different burstiness, interarrival times, and parallelism~\cite{versluis2019tracearchive}. An effective benchmarking framework should allow for generating workloads that reflect these characteristics rather than a `one-size-fits-all' approach. Current benchmarking often overfits well-known publicly available workload traces due to the scarcity of realistic workloads available~\cite{amvrosiadis2018workloaddiveristy}. 
    Realism is essential for effectively testing and improving AI agents, as evidenced by practitioners at CompanyX who faced delays in AI agent deployment due to the difficulty in obtaining genuine user interaction and traffic patterns.

    \item \label{req:8} \textbf{\textit{Coverage of the operations lifecycle.}}
    % 
    % \textit{It must allow different stages of the operations and reliability lifecycle to be incorporated, including fault and anomaly detection, root cause analysis, recovery, mitigation and prevention.}
    % 
    The incident management process can have diverse goals and scopes, and the framework should support the different stages that correspond to them, like fault detection, root-cause analysis, and mitigation. It should allow proactive strategies to predict and preempt failures and reactive strategies to detect and correct errors.

    \item \label{req:9} \textbf{\textit{Sufficient Observability into applications.}}
    % 
    % \textit{There should be sufficient observability into various layers to help assist with different types of operations.}
    % 
    Adequate visibility into the different aspects of the system and application environment should be available to detect faults and their impact. It includes interfaces to access logs, telemetry, traces, KPIs, and documentation such as incident reports, standard procedures, etc. Tools to explore configuration files and source code may be beneficial in acquiring the necessary context for localization and mitigation.

    \item \label{req:10} \textbf{\textit{Adequate controls for agent operations.}}
    % 
    % \textit{There should be sufficient controls into various layers to help assist with the operations.}
    % 
    Fault correction may require modifying configuration files or restarting VMs or services. Even fault detection may require the agent to run test suites to gain more visibility into the possible cause. The framework should enable such actions by having hooks into different layers.
    % into different ecosystem layers.

    % 7 -> 8, 8 -> 9, 9 -> 10

\end{enumerate}

\subsection{Design Decisions}
\label{sec:design}
% \own{MS/GS}

To address the requirements outlined in ~\cref{sec:requirements} (annotated below), we recommend several key design choices to build \ourbench{}, a new standardized evaluation framework for AIOps:

\subsubsection{\textbf{Bring your Services Alive}}\label{sec:d1} 
We propose evaluating with live services, that are designed and implemented using different architectures at various scales \ref{req:3}, and capable of operating in various contexts, from sandbox to production \ref{req:5}. Real-world cloud services present complex behaviours, challenging AIOps agents with variability in workload patterns, diversity of faults, and intricacies of service dependencies. This approach closely mirrors the operational and reliability challenges faced in production systems, making the benchmark directly applicable to practitioners' needs.
Further, we choose automation tools like \code{Helm}~\cite{helm} and \textsc{Blueprint}~\cite{blueprint} for repeatable and consistent setups \ref{req:4} and interfaces to extend to new services, preserving its applicability and rigour \ref{req:1}.
This approach provides a comprehensive, realistic, and relevant evaluation platform to advance AIOps research and practice.

\subsubsection{\textbf{Real Faults, Real Challenges}}\label{sec:d2} 
We recommend incorporating dynamic workload and fault generators to simulate real-world conditions accurately. These generators are designed to produce a wide range of realistic traffic patterns and operational challenges \ref{req:7}, from typical user behaviour to peak loads and various fault scenarios, including kernel failures, network issues, and configuration bugs \ref{req:6}.
This approach not only tests adaptability and robustness but also mitigates the risk of ``training data contamination'' for LLM-powered tools. A key implication of this choice is that the state of a faulty service (say telemetry or logs) is only relevant and observed in the context of an injected fault -- making it publicly unavailable to seep into training data.

\subsubsection{\textbf{Evaluate Across the Board}}\label{sec:d3} 
The combination of live services (\S~\ref{sec:d1}) and real faults (\S~\ref{sec:d2}) allow us to reliably evaluate the impact of agents on various AIOps tasks and performance dimensions.
Here, one can use generated faults independently or in conjunction to create benchmark problems for agents. Notably, one can use a single fault (e.g., network misconfiguration) to evaluate tasks across the board -- from detection to resolution \ref{req:8}.
Evaluation involves quantitative dimensions, such as performance, time, resource usage, dollar cost, and other metrics beyond accuracy~\cite{agentsthatmatter}.
Furthermore, to reliably compare agents, we emphasize qualitative evaluation of their traces by a human or LLM-as-a-Judge \cite{llmjudge}.

\subsubsection{\textbf{Orchestrate the Agent-Cloud-Interface}}\label{sec:d4} 
Typically, service engineers operate cloud environments with various programming (e.g. APIs, database queries) and user interfaces (incident portals). 
% Naturally, existing interfaces to the cloud have been designed with only a human user in mind.
% % 
% Retrofitting existing interfaces to LLMs and agents can be suboptimal. 
Existing UI to the cloud are designed only for a human,
and not amenable for LLMs and agents.
E.g., humans reliably ignore extra information while the same can distract the context and harm performance for agents~\cite{litm}.
Inspired by human-computer interaction (HCI), \citet{sweagent} introduce \textit{agent-computer-interface}, finding that agents can similarly benefit from better-designed interfaces for coding tasks.
We posit the same for AIOps and envision an \textbf{\textit{Agent-Cloud-Interface} (ACI)}. 
% As shown in \cref{fig:overview}, 
The ACI is an Orchestrator between the agent and the cloud (\cref{fig:overview}) which specifies both the actions available and how the service's state is conveyed back to the agent as the observation of its actions \ref{req:2}.
It simplifies the action space into a concise list of APIs, each documented to ensure that agents can make meaningful progress towards objectives \ref{req:10}. Also, the Orchestrator takes actions on behalf of the agent and returns high-quality feedback (e.g., outputs, error messages, etc.).

\subsubsection{\textbf{Abstract Environments, not agents}}\label{sec:d5} 
Two sides to a real-time evaluation are the agent and the environment. For AIOps, the agent can be a DevOps engineer or an AIOps tool (e.g., LLM-powered agent). The environment is a deployed application (e.g., \code{SocialNetwork}) that the user interacts with. Here, we suggest providing abstractions for the environment, not the agent. This choice maximizes flexibility in implementing and evaluating various kinds of tools and agents \ref{req:2}. Consequently, \ourbench{} provides sufficient information (task description, available APIs/actions, and additional instructions) to solve a problem in the benchmark.

\subsubsection{\textbf{Observe Everything, Everywhere, All at Once.}}\label{sec:d6} 
Observability captures a system's internal states from its external outputs. It traditionally includes (1) \textit{traces} detailing the end-to-end request paths through distributed entities; (2) \textit{application logs} as textual records of runtime operations; and (3) \textit{metrics} monitoring component health.
We suggest an observability layer to collect not only canonical telemetry data but also other system indicators, such as cluster information, including logs, running status, and configurations \ref{req:9}.
But increased observability 
% offers detailed system status, it 
can overwhelm AIOps tools with data volume and complex data types. Therefore, we offer flexible APIs for users to select the specific information they need, ensuring tailored and comprehensive observability.

\Paragraph{Summary}
In summary, these design decisions prioritize creating a framework that is:
\begin{enumerate}[leftmargin=*]
    \item \textbf{Realistic}: By using live services, workloads, and faults that mirror real-world operational challenges.
    \item \textbf{Scalable}: Through dynamic workload and fault generators that can create new problem scenarios at varying scales.
    \item \textbf{Reliable}: Evaluating tasks and performance dimension across the operations lifecycle.
    \item \textbf{Observable}: Through rich telemetry data and actual service interactions, ensuring reliable evaluation.
    \item \textbf{Flexible}: With the Agent-Cloud-Interface (ACI) that supports plugging a diverse range of tools.
    \item \textbf{Extensible}: With the ability to easily incorporate new services, workloads, and faults.
\end{enumerate}

% Applications, Fault generation, Fault Mitigation
% Expose APIs/interfaces for each of these. 
% Apps FW needs to expose APIs to receive workload/requests, APIs for observability, APIs for fault injection and APIs for actions to mitigate.
% Fault generation APIs: exposed at the application layer, system layer. Chaos engg can use this API to generate fault workloads.
% AIOps APIs: Make use of the App observability and App action APIs to identify and solve the problem. Even a human tester can do it.
% Observability should also expose details for Success Metrics?
% Extensible. Reproducibile. What if?
% Agent--Cloud Interface: Cloud spans the infra, platform and software layers.

\section{System Architecture}
% \pb{0.75pgs}
% \own{GS/MS + MH/YC}
\label{sec:arch}
This section expands on how the decisions discussed in \S~\ref{sec:design} lead to an prototype system architecture for \ourbench{}.
\cref{fig:overview} shows our system architecture consisting of $5$ key pieces.

\subsection{\textbf{Orchestrator}}

\ourbench{} strictly separates the Agent and the Application Service using an intermediate Orchestrator. It provides several interfaces for other system parts to integrate and extend.
% Agent-Orch
First, it establishes a \code{session} with an Agent to share information about benchmark problems: (1) the problem description, (2) instructions (e.g., response format), and (3) available APIs to call as actions.
As shown in \cref{fig:detect}, the APIs are a set of documented tools, e.g., \code{get_logs}, \code{get_metrics}, \code{exec_shell},
designed to help the Agent solve a task.
There are no restrictions on the Agent's implementation; the Orchestrator poses problems and polls it for the next action to perform given the previous result. Each action must be a valid API call, which the Orchestrator validates and carries out.
% Serv-Orch
The Orchestrator has privileged access to the deployment and can take arbitrary actions (e.g., scale-up, redeploy) using appropriate tools (e.g., \code{helm}, \code{kubectl}) to resolve problems on behalf of the Agent.
% Generators-Orch
Lastly, the Orchestrator calls workload and fault generators to create service disruptions, which serve as live benchmark problems. \ourbench{} provides additional APIs to extend to new services and generators.

\subsection{\textbf{Service}}
\ourbench{} abstracts a diverse set of services to reflect the variance in production environments. This includes live, running services implemented using various architectural principles, including microservices, serverless and monolithic.
We also leverage open-sourced application suites such as \textsc{DeathStarBench}~\cite{deathstarbench} as they provide artifacts, like source code and commit history, along with run-time telemetry. Adding tools like \textsc{BluePrint}~\cite{blueprint} can help scale to other academic~\cite{trainticket,teastore,serverlessbench,sriraman2018mu, mubench,xfbench-2024} and production services~\cite{overleaf} and also seamlessly deploy new variants of these services.

\subsection{\textbf{Workload Generator}}
The workload generator in \ourbench{} plays a crucial role by creating simulations of both faulty and normal scenarios. It receives specifications from the Orchestrator, such as the task, desired effects, scale, and duration. 
The generator can utilize a model trained on real production traces, to generate workloads that align with these specifications.
Faulty scenarios may simulate conditions like resource exhaustion, exploit edge cases, or trigger cascading failures, inspired by real incidents. Normal scenarios mimic typical production patterns, such as daily activity cycles and multi-user interactions. 
When various characteristics (e.g., service calls, user distribution, arrival times) can lead to the desired effect, multiple workloads can be stored in the problem cache for use by the Orchestrator. In coordination with the Fault Generator (~\ref{subsection:faultgen}), the workload generator can also create complex fault scenarios with workloads.
%(e.g., high load with mild resource contention). 

\subsection{\textbf{Fault Generator}}
\label{subsection:faultgen}
\ourbench{} has a novel push-button fault generator designed for generic applicability across various cloud scenarios. Our approach integrates application and domain knowledge to create adaptable policies and ``oracles'' compatible with AIOps scenarios.
This includes fine-grained fault injection capable of simulating complex failures inspired by production incidents. 
Additionally, it can inject faults at various system levels, exposing root causes while maintaining semantic integrity and considering interdependencies between cloud microservices. The fault injector's versatility can enhance the reliability and robustness of cloud systems by enabling thorough testing and evaluation of AIOps capabilities.

\subsection{\textbf{Observability}}
\ourbench{} is equipped with an extensible observability layer designed to provide comprehensive monitoring capabilities across various system layers for any AIOps tool. \ourbench{} collects a wide array of telemetry data, including (1) traces from Jaeger detailing the end-to-end paths of requests through distributed systems, (2) application logs formatted and recorded by Filebeat and Logstash, and (3) system metrics monitored by Prometheus. Additionally, \ourbench{} also captures lower-level system information such as \code{syscall} logs and cluster information.
% encompassing running statuses and configurations. 
As mentioned, we handle the potential data overload through
% and complexity, \ourbench{} provides 
flexible APIs to tune the 
% , allowing users to selectively access the specific data they need and collect the 
telemetry data relevant to the AIOps tools.

% Figure{agent Trace}
\begin{figure}[t]
    \centering
    \input{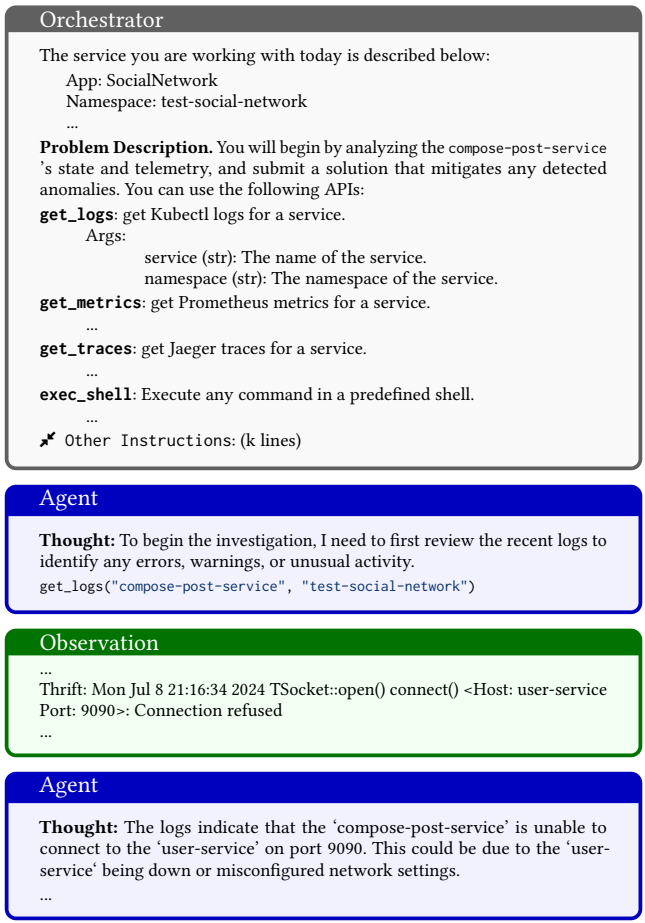}
    \caption{Orchestrator shares a problem with APIs to be used as actions and an agent detects an anomaly in the logs.}
    \label{fig:detect}
    %\vspace{-0.05in}
\end{figure}

\begin{figure}[t]
    \centering
    \input{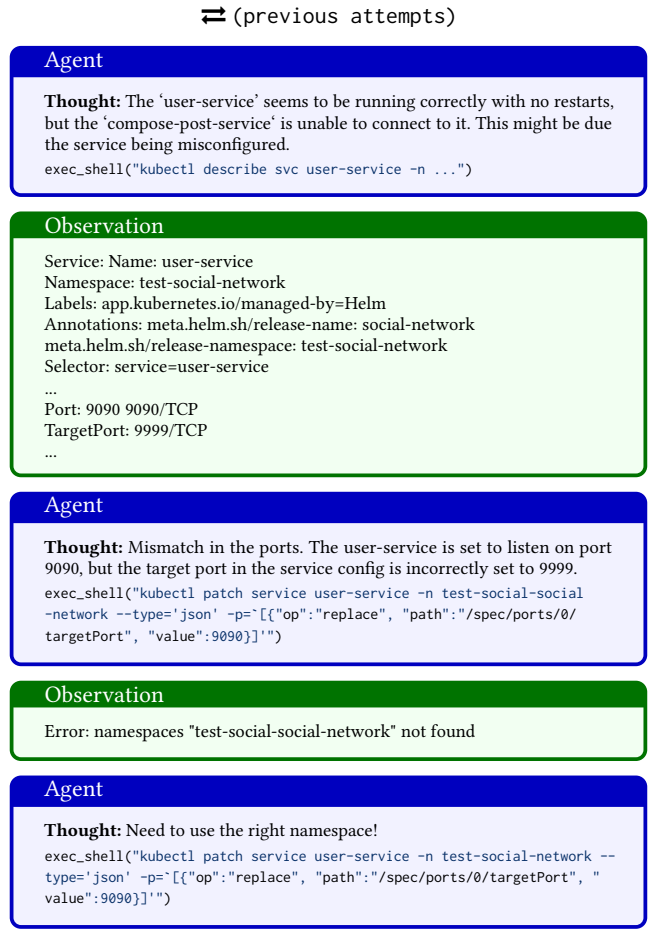}
    \caption{Agent identifying the root-cause as a K8S misconfiguration fault. It then goes about mitigation by patching it.}
    \label{fig:mitigate}
    %\vspace{-0.05in}
\end{figure}

\section{Case Study and Insights}
% \pb{0.75pgs}
% \own{MS}
\label{sec:eval}

As a proof-of-concept, we implement a prototype of \ourbench{} and evaluate an LLM agent on an AIOps incident mitigation task. Here, we aim to demonstrate the potential of \ourbench{} to standardize evaluation for AIOps tools and uncover novel insights.

\Paragraph{Setup} We deploy the \code{SocialNetwork} application from \textsc{DeathStarBench} \cite{deathstarbench} on a Kubernetes cluster. We instantiate the fault generator to induce a realistic misconfiguration fault at the virtualization layer: the target port of a microservice is misconfigured, causing connectivity issues with other microservices. We generate traffic by using an exponential workload pattern from the \code{wrk} tool.
We study a \textsc{ReAct} \cite{react} agent with a \textsc{GPT-4} \cite{gpt4} backend. \textsc{ReAct} is a popular LLM-agent paradigm that leverages interleaved Thought-Action traces, utilizing LLMs' reasoning and planning abilities.

\Paragraph{Evaluation Goals}
% R4 - quant eval (Success, TTD, TTM) and qual (Human analysis of the trace)
% R7 - evaluating multiple stages: detect, RCA, and mitigate
% R8 - observability at application-level
% R9 - 3 read APIs (e.g., k8s pod logs) and 1 read/write API (exec_shell).
% 
This case study aims to showcase several of the requirements in \cref{sec:requirements}.
We address \ref{req:4} by employing quantitative metrics (Success, Time-to-Detect (TTD), Time-to-Mitigate (TTM), and Efficiency) and qualitative analysis of the agent's actions.
We demonstrate \ref{req:8} by covering multiple stages: detection, root-cause analysis, and mitigation.
% .
\ref{req:9} is ensured through application observability (logs, metrics, traces), and the well-defined APIs for the agent, including a \texttt{shell}, fulfill \ref{req:10}.

\Paragraph{Key Insights} \cref{fig:detect} illustrates the agent's trace while attempting to first detect a service anomaly. It then takes a series of actions, as shown in \cref{fig:mitigate}, to identify the root cause and mitigate the fault.
Overall, it successfully detected the problem in $14$ secs (TTD) and mitigated it in $36$ secs (TTM). We measured the average resource efficiency, and found that it takes $10$--$12$ interactions costing around \$$0.25$
Below, we distill key insights from this case study:

\HighlightPara{\circled{1} Importance of Observability}
The agent's ability to detect the misconfiguration fault hinged on the detailed telemetry data provided by \ourbench{}. For example, the agent identified repeated connection refusals in the logs, which made it hypothesize a potential misconfiguration. This highlights the critical role observability will play in AIOps in the future.

\HighlightPara{\circled{2} AIOps needs Efficient Actions} We found that the efficiency of the available actions influenced the agent's performance. For instance, too few APIs limited its ability to explore solutions, while too many arguments for each API hindered its performance. Also, flexible APIs (like \code{exec_shell}) were pivotal to the agent balancing explore and exploit actions. This reiterates the importance of a well-designed Agent-Cloud-Interface (ACI).

\HighlightPara{\circled{3} Fault Injection Reflects Real-World Complexity} Interestingly, injecting simple faults (like a K8s misconfiguration) into a real service proves challenging problems for advanced models like \textsc{GPT-4} \cite{gpt4}, requiring 10+ interaction rounds to reach a solution. This demonstrates how automated fault generators could accelerate the creation of rigorous testbeds for reliable AIOps evaluation.

\HighlightPara{\circled{4} Incorporating Service Dependency Tools} The trace in \cref{fig:detect} demonstrates the agent's implicit understanding of service dependencies just via logs. While promising, for complex applications, the agent could spend significant time traversing the call graph with only partial views of the system \cite{fogofwar}.
% \ms{cite fog-of-war}
% 
Concurrent efforts in coding agents have shown the power of code dependency analysis for repository-level tasks \cite{r2e, autocoderover, aider}. We believe future research can similarly look at augmenting AIOps agents with tools for explicit service dependencies and impact analysis.

\HighlightPara{\circled{5} Error Handling goes a long way} In \cref{fig:mitigate}, when the agent encountered a syntax error, it quickly corrected the mistake and retried the command. We even find that poor error messages hinder the agent's performance. This insight emphasizes the need for robust error propagation for actions throughout the operations and reliability lifecycle of systems.

% \ms{todo: conclude motivating how such a framework could uncover many more insights.}
\section{Conclusion}
\pb{0.25pgs}
% \own{YS}
\label{sec:conclude}
% \todo{}
In this vision paper, we have framed the requirements and design principles for a framework to build, test, compare and improve Agents used for the operational management of cloud services. We also discuss our prototype implementation, \ourbench{}, and report preliminary results. This forms the kernel for a more comprehensive framework that will address the key gap in standardized framework for building agents to help achieve autonomous clouds.

%\clearpage
\balance
\bibliographystyle{ACM-Reference-Format}
\bibliography{bibliography}
\end{document}